\documentclass{inprep}
\usepackage{graphicx}

\newcommand{\ch}{\mathop{\rm ch}\nolimits}
\newcommand{\sh}{\mathop{\rm sh}\nolimits}
\renewcommand{\tanh}{\mathop{\rm th}\nolimits}
\newcommand{\va}[1]{\left<#1\right>}

\begin{document}

\begin{center}
\Large Budker Institute of Nuclear Physics
\\[35mm]
\large V.~N.~Baier and A.~G.~Grozin
\\[10mm]
\large Sum rules for $1/m$ corrections
\\[3mm]
\large to the form factors of semileptonic $B$ decays
\\[20mm]
\Large Budker INP 92-80
\\
\vfill
NOVOSIBIRSK
\\[2mm]
1992
\end{center}
\thispagestyle{empty}
\newpage

\begin{center}
\bf Sum rules for $1/m$ corrections
\\
\bf to the form factors of semileptonic $B$ decays
\\[3mm]
\it V.~N.~Baier$^1$ and A.~G.~Grozin$^{2,3}$
\\[3mm]
\rm Budker Institute of Nuclear Physics,
\\
630090 Novosibirsk 90, Russia
\\[5mm]
ABSTRACT
\end{center}
\begin{quotation}
Semileptonic $B$ decays are described by the Isgur-Wise form factor
to the leading order in $1/m$;
four new functions appear in the first order~\cite{Luke}.
Values of these functions are crucial for the applicability
of the whole approach to processes involving $c$ quark.
We obtain the sum rules for three subleading form factors from
the QCD sum rules with finite masses by expanding to the first order in $1/m$.
The results respect the pattern of the first $1/m$ corrections
established in HQET, and obey the Luke's theorem.
The numerical estimates show that $1/m_c$ corrections are sizable
but not catastrophic.
\end{quotation}
\vfill
\begin{flushright}
\copyright{} Budker Institute of Nuclear Physics
\end{flushright}
\addtocounter{footnote}{1}
\footnotetext{Internet address: \tt BAIER@INP.NSK.SU}
\addtocounter{footnote}{1}
\footnotetext{Supported in part by grant from the Soros fund}
\addtocounter{footnote}{1}
\footnotetext{Internet address: \tt GROZIN@INP.NSK.SU}
\thispagestyle{empty}
\newpage

{
\parindent=0pt
\begin{tabular}{r}
\hline
\hspace{112mm}
\end{tabular}
\par
}

\vspace{50mm}

\section{Introduction}
\label{Intro}

Recently a significant progress has been achieved in the heavy quark physics
in the framework of the Heavy Quark Effective Theory (HQET)~\cite{HQET},
see also the reviews~\cite{HQETrev} and references cited therein.
The matrix elements of the vector current $V_\mu=\overline{c}\gamma_\mu b$
for $B\to D^{(*)}$ decays
\begin{eqnarray}
\va{D|V_\mu|B} &=& \sqrt{m_B m_D}
\left(\xi_+(v+v')_\mu + \xi_-(v-v')_\mu\right),
\nonumber\\
\va{D^*|V_\mu|B} &=& \sqrt{m_B m_{D^*}} \xi_V
i\varepsilon_{\mu\nu\alpha\beta} e^*_\nu v'_\alpha v_\beta,
\label{ff}
\end{eqnarray}
to the leading order in $1/m$ are expressed via the Isgur-Wise form factor:
$\xi_+=\xi_V=\xi(\ch\varphi)$, $\xi_-=0$, where $\ch\varphi=vv'$,
$v$, $v'$ are 4-velocities of $B$, $D^{(*)}$, $\xi(1)=1$~\cite{IW,FGGW}.
First $1/m_c$ corrections involve four new functions~\cite{Luke};
$1/m_b$ corrections contain no new elements~\cite{NR}:
\begin{eqnarray}
\xi_+ &=& \xi \left[ 1 + \left(\frac1{m_c}+\frac1{m_b}\right)\rho_1 \right],
\nonumber\\
\xi_- &=& \xi \left(\frac1{m_c}-\frac1{m_b}\right)
\left(-\frac{\varepsilon}{2}+\rho_4\right),
\label{xi}\\
\xi_V &=& \xi \left[ 1
+ \left(\frac1{m_c}+\frac1{m_b}\right)\frac{\varepsilon}{2}
+ \frac{\rho_2}{m_c} + \frac{\rho_1-\rho_4}{m_b} \right],
\nonumber
\end{eqnarray}
where the HQET ground state meson's energy
$\varepsilon=m_B-m_b=m_D-m_c=m_{D^*}-m_c$
(these differences are equal up to $1/m$ corrections).
The Luke's theorem~[5--7] states that $\rho_1(1)=0$, $\rho_2(1)=0$.

The most interesting physical applications of HQET
are those to $b\to c$ weak transitions.
Therefore the applicability of the whole approach
to processes involving $c$ quark (i.~e.\ the size of $1/m_c$ corrections)
is a crucial question of HQET.
If these corrections are large, HQET has only a purely academic interest;
if they are modest, it can be applied to real-world processes.

A nonperturbative method is needed to calculate form factors.
QCD sum rules~\cite{SVZ,ISNR} were used for investigation
of the form factors~(\ref{ff}) at finite $m_b$, $m_c$ in~\cite{BG,OS}.
HQET sum rule for the Isgur-Wise form factor was considered in~\cite{R,N}.
It coincides with the limit $m_{b,c}\to\infty$ of the QCD sum rules.
Results for finite and infinite masses are compared in~\cite{Ball}.
There are two alternative ways to obtain sum rules
for the subleading form factors $\rho_i$.
One can expand the known finite-mass QCD results to the first order in $1/m$.
Alternatively, one can start from the HQET lagrangian and currents
in the first order in $1/m$.
The second way gives more insight into the sources
of the heavy quark symmetry breaking
(e.~g.\ the heavy quark chromomagnetic moment vertex);
the general theorems of HQET (like~(\ref{xi}) and the Luke's theorem)
can be traced in the calculation.
But the first way is less labour-consuming provided that the QCD results
are already known, and allows to consider easily also higher $1/m$ corrections.
It also gives a strong check of the QCD results.
Here we use this way.

The situation is similar in a simpler case of 2-point sum rules.
Here the QCD (Borel-transformed) sum rules~[15--19]
coincide with the leading-order HQET sum rules~\cite{Shuryak,BG2}
in the limit $m_{b,c}\to\infty$.
The first $1/m$ correction was obtained by expanding
the QCD results~\cite{DPES},and also in the framework of HQET~\cite{N2}.

\section{Sum rules}
\label{Sr}

We consider three-point correlators
\begin{eqnarray}
K_\mu(p_b,p_c) &=& \int dx_b dx_c e^{-ip_b x_b+ip_c x_c}
\va{T j_B(x_b) V_\mu(0) j^+_D(x_c)}
\nonumber\\
&=& K_+(p_b^2,p_c^2,t) p_\mu + K_-(p_b^2,p_c^2,t) q_\mu,
\label{corr}\\
K_{\mu\nu}(p_b,p_c) &=& K_V(p_b^2,p_c^2,t)
i\varepsilon_{\mu\nu\alpha\beta} p_\alpha q_\beta,
\nonumber
\end{eqnarray}
where $K_{\mu\nu}$ is similar to $K_\mu$ with $J_{D^*\nu}$ instead of $j_D$;
$j_B=\overline{q}\gamma_5 b$, $j_D=\overline{q}\gamma_5 c$,
$j_{D^*\nu}=\overline{q}\gamma_\nu c$ are the currents
with the quantum numbers of $B$, $D$, $D^*$;
$p=p_b+p_c$. $q=p_b-p_c$, $t=q^2$.
We have calculated perturbative spectral densities $\rho_i(s_b,s_c,t)$
and quark condensates' contributions $K^q_i(p_b^2,p_c^2,t)$ up to dimension 6
for the invariant functions $K_i(p_b^2,p_c^2,t)$ using REDUCE~\cite{REDUCE};
part of the results was published in~\cite{BG}.

In order to consider the limit $m_{b,c}\to\infty$,
we proceed to the new variables $p_{b,c}^2=m_{b,c}^2+2m_{b,c}\omega_{b,c}$,
$t=m_b^2+m_c^2-2m_b m_c\ch\varphi$.
In these variables the support of perturbative spectral densities~\cite{BG}
is the wedge
\begin{equation}
e^{-\varphi} < \omega_b/\omega_c < e^\varphi.
\label{wedge}
\end{equation}
At $t>0$ spectral densities are singular at the parabola
$s_b^2+s_c^2+t^2-2s_b s_c-2s_b t-2s_c t=0$
which touches the boundary of the physical region~(\ref{wedge}) at
\begin{equation}
\omega^*_b = \frac{m_b m_c\sh\varphi}{t} \left(m_b-m_c e^{-\varphi}\right),
\quad
\omega^*_b = \frac{m_b m_c\sh\varphi}{t} \left(-m_c+m_b e^\varphi\right).
\label{touch}
\end{equation}
The double dispersion representation should be modified
above this point~\cite{BBD}.
In the limit $m_{b,c}\to\infty$, this area goes to infinity and hence
lies outside the lowest mesons' duality region essential for the sum rules.

Making the double Borel transform from the variables $\omega_{b,c}$
to $E_{b,c}$, we obtain QCD sum rules (for finite $m_{b,c}$)
\begin{eqnarray}
&&\frac{f_B m_B^2}{m_b} \frac{f_D m_D^2}{m_c} \xi_\pm(\ch\varphi)
e^{-(\varepsilon_B+\varepsilon_D)/(2E)}
= \frac{2m_b m_c}{\sqrt{m_B m_D}}
\nonumber\\
&&\quad\biggl\{ \int \left[(m_B\pm m_D)\rho_+ + (m_B\mp m_D)\rho_-\right]
e^{-(\omega_b+\omega_c)/(2E)} d\omega_b d\omega_c
\nonumber\\
&&\quad{} + 4E^2\hat{B} \left[(m_B\pm m_D)K^q_+ + (m_B\mp m_D)K^q_-\right]
\biggr\},
\label{QCD}\\
&&\frac{f_B m_B^2}{m_b} f_{D^*}m_{D^*} \xi_V(\ch\varphi)
e^{-(\varepsilon_B+\varepsilon_{D^*})/(2E)}
\nonumber\\
&&\quad{} = 8m_b m_c \sqrt{m_B m_{D^*}}
\left\{ \int \rho_V e^{-(\omega_b+\omega_c)/(2E)} d\omega_b d\omega_c
+ 4E^2 \hat{B} K^q_V \right\},
\nonumber
\end{eqnarray}
where we have taken $E_b=E_c=2E$;
$m_B^2=m_b^2+2m_b\varepsilon_B$, and similarly for $D^{(*)}$.
The integrals are calculated over the part of the wedge~(\ref{wedge})
dual to the lowest mesons in both channels,
i.~e.\ the wedge minus the higher states' continuum region.
It is convenient to introduce the variables $\omega$, $\eta$
instead of $\omega_{b,c}=\omega\left(1\pm\eta\tanh^2\frac{\varphi}{2}\right)$
so that the physical region~(\ref{wedge}) is $-1<\eta<1$,
and to measure quantities with the dimension of energy (such as $E$)
in units of $k$, $k^3=\frac{\pi^2}{6}\left|\va{\overline{q}q}\right|$
($m_0=4kE_0$).
Expanding to the first order in $1/m_{b,c}$, we obtain
\begin{eqnarray}
&&\frac{f_B m_B^2 f_D m_D^2}{(m_b m_c)^{3/2}|\va{\overline{q}q}|}
e^{-(\varepsilon_B+\varepsilon_D)/(2E)} \xi_+(\ch\varphi)
\nonumber\\
&&\quad{} = A(\varphi,E)
+ \left(\frac1{m_c}+\frac1{m_b}\right) k B_1(\varphi,E),
\nonumber\\
&&\frac{f_B m_B^2 f_D m_D^2}{(m_b m_c)^{3/2}|\va{\overline{q}q}|}
e^{-(\varepsilon_B+\varepsilon_D)/(2E)} \xi_-(\ch\varphi)
\nonumber\\
&&\quad{} = \left(\frac1{m_c}-\frac1{m_b}\right)
\left[ -\frac{\varepsilon}{2}A(\varphi,E) + k B_4(\varphi,E) \right],
\label{SR3}\\
&&\frac{f_B m_B^2 f_{D^*} m_{D^*}}{m_b^{3/2}m_c^{1/2}|\va{\overline{q}q}|}
e^{-(\varepsilon_B+\varepsilon_D)/(2E)} \xi_V(\ch\varphi)
\nonumber\\
&&\quad{} = \left[ 1
+ \left(\frac1{m_c}+\frac1{m_b}\right)\frac{\varepsilon}{2} \right]A(\varphi,E)
\nonumber\\
&&\quad{} + \frac{kB_2(\varphi,E)}{m_c}
+ \frac{k\left(B_1(\varphi,E)-B_4(\varphi,E)\right)}{m_b}.
\nonumber
\end{eqnarray}
Here
\begin{eqnarray}
&&A(\varphi,E) = \frac1{4\ch^4\frac{\varphi}2}
\int \omega^2 e^{-\omega/E} d\eta d\omega
+ 1 - \frac{2\ch\varphi+1}{3} \frac{E_0^2}{E^2}
+ \frac{\alpha_s\ch\varphi}{27\pi E^3},
\nonumber\\
&&B_1(\varphi,E) = \frac1{16\ch^6\frac{\varphi}2}
\int \left(\ch\varphi - 5 - \eta^2\ch\varphi + \eta^2\right)
\omega^3 e^{-\omega/E} d\eta d\omega
\nonumber\\
&&\quad{} + 2\frac{E_0^2}{E}
+ \frac{\alpha_s(4\ch\varphi-1)}{27\pi E^2},
\label{AB}\\
&&B_2(\varphi,E) = -\frac1{8\ch^6\frac{\varphi}2}
\int (3-\eta^2) \omega^3 e^{-\omega/E} d\eta d\omega
- \frac{8\alpha_s}{27\pi E^2},
\nonumber\\
&&B_3(\varphi,E) = \frac1{8\ch^4\frac{\varphi}2}
\int (1-\eta^2) \omega^3 e^{-\omega/E} d\eta d\omega
+ \frac{2E_0^2}{3E} + \frac{\alpha_s(4\ch\varphi+1)}{27\pi E^2}.
\nonumber
\end{eqnarray}
Similar expansion of 2-point sum rules gives
\begin{eqnarray}
&&\frac{f_D^2 m_D^4}{m_c^3|\va{\overline{q}q}|} e^{-\varepsilon_D/E}
= A(0,E) + \frac{2kB_1(0,E)}{m_c},
\label{SR2}\\
&&\frac{f_{D^*}^2 m_{D^*}^2}{m_c|\va{\overline{q}q}|} e^{-\varepsilon_{D^*}/E}
= A(0,E) + \frac{2kB_2(0,E)}{m_c}
\nonumber
\end{eqnarray}
(the sum rule for $f_B$ is similar to the one for $f_D$).
From~(\ref{SR3}) and~(\ref{SR2}) we obtain the sum rules
for $\xi(\ch\varphi)$~\cite{R,N} and for $\rho_{1,2,4}(\ch\varphi)$
\begin{eqnarray}
&&\xi(\ch\varphi) = \frac{A(\varphi,E)}{A(0,E)},
\quad
\rho_1(\ch\varphi)/k
= \frac{B_1(\varphi,E)}{A(\varphi,E)} - \frac{B_1(0,E)}{A(0,E)},
\label{SR}\\
&&\rho_1(\ch\varphi)/k
= \frac{B_2(\varphi,E)}{A(\varphi,E)} - \frac{B_2(0,E)}{A(0,E)},
\quad
\rho_4(\ch\varphi)/k = \frac{B_4(\varphi,E)}{A(\varphi,E)}.
\nonumber
\end{eqnarray}

The results of $1/m$ expansion of the QCD sum rules
obey the structure~(\ref{xi}) and the Luke's theorem.
It is easy to explain why $\rho_1(1)=0$.
At finite $m_b=m_c$, $q\to0$ the 3-point correlator
is related to the 2-point one by the Ward identity
$K_+(p^2,p^2)=\frac{d\Pi(p^2)}{dp^2}$~\cite{BG},
and hence the QCD sum rule reproduces the exact result $\xi_+(t=0)=1$.
At infinite $m_{b,c}$, the Ward identity is
$K(\omega,\omega)=\frac{d\Pi(\omega)}{d\omega}$,
and hence the HQET sum rule reproduces the exact result $\xi(1)=1$.
Then we obtain from~(\ref{xi}) $\rho_1(1)=0$.
Therefore it is not accidental that the $1/m$ correction
to the sum rule for $f_D$~(\ref{SR2}) is described by the same function $B_1$
as to the sum rule for $\xi_+$~(\ref{SR3}) but at $\varphi=0$.
The similar fact for the sum rule for $f_{D^*}$~(\ref{SR2})
and $1/m_c$ correction to $\xi_V$~(\ref{SR3}) looks like a miracle
in the framework of QCD sum rules, but it leads to the Luke's theorem
for $\rho_2$ in~(\ref{SR}).
The dependence between $1/m_b$ correction to $\xi_V$
and $1/m$ corrections to $\xi_\pm$ in~(\ref{SR3}) is one more miracle.
From~(\ref{SR2}) we obtain (compare with~\cite{DPES})
\begin{equation}
f_D = \frac{\tilde{f}}{\sqrt{m_c}} \left(1 + \frac{c_1}{m_c}\right),
\quad
f_{D^*} = \frac{\tilde{f}}{\sqrt{m_c}} \left(1 + \frac{c_2}{m_c}\right),
\label{fm}
\end{equation}
where the coefficients are given by the sum rules
\begin{eqnarray}
&&\frac{\tilde{f}^2}{|\va{\overline{q}q}|} e^{-\varepsilon/E} = A(0,E),
\label{SRf}\\
&&c_1/k = \frac{B_1(0,E)}{A(0,E)} - 2\varepsilon/k,
\quad
c_2/k = \frac{B_2(0,E)}{A(0,E)} - \varepsilon/k.
\end{eqnarray}

\section{Results}
\label{Res}

At the standard values of the condensates, $k=260$--280MeV,
$E_0=\frac{m_0}{4k}=0.85$.
The 2-point sum rule~(\ref{SRf}) for $\tilde{f}^2$
was analyzed in~\cite{Shuryak}.
The continuum threshold is $\omega_0=3$
and the meson energy is $\varepsilon=1.65$
in the Borel parameter plato $E=1.7$--2.5 (all in $k$ units).
Now we know that the perturbative correction to these results
is large~\cite{BG2}.
But this correction is unknown in the 3-point sum rule.
We may hope that it will be partially compensated
in the ratio of 3-point to 2-point sum rules.
For consistency, we use the 2-point results without perturbative corrections
as an input for the 3-point sum rules analysis.
In any case, even a rough estimate of $\rho_i$ is currently valuable.

\begin{figure}[ht]
\begin{center}
\includegraphics{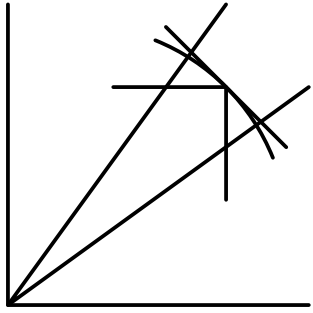}
\end{center}
\caption{Continuum models}
\label{F1}
\end{figure}

In the case of 3-point sum rules, e.~g.\ for $\xi(\ch\varphi)$~(\ref{SR}),
the higher states' continuum spectral density is modelled
by the perturbative spectral density with some smooth curve
as a continuum threshold (Fig.~\ref{F1}).
At $\varphi\to0$ the wedge becomes infinitesimally narrow,
and the continuum starts at $\omega_0$ for consistency
with the 2-point sum rules.
The wedge area is $\sim\varphi$,
therefore the spectral density must be $\sim1/\varphi$
in order to yield the perturbative contribution $\sim1$
(of course, the explicit calculations confirm it).
The simplest continuum model is that with the straight line threshold.
If one chooses any smooth curve instead,
the difference area is $\sim\varphi^3$ (Fig.~\ref{F1}),
and the variation of $\xi(\ch\varphi)$ is $\sim\varphi^2$.
It influences the slope of $\xi(\ch\varphi)$ at 1;
this freedom is analogous to the freedom of choosing the continuum threshold
in the 2-point sum rule~(\ref{SRf}) for $\tilde{f}^2$.
But if one chooses a continuum threshold with a cusp in the physical region
(as was done in~\cite{R}), the difference area is $\sim\varphi^2$
(Fig.~\ref{F1}), and the variation of $\xi(\ch\varphi)$ is $\sim\varphi$.
This gives an infinite slope of $\xi(\ch\varphi)$ at 1,
what contradicts to its general analytical properties.
Besides that, there are no physical reasons
for the continuum threshold to be non-smooth.
Here we restrict ourselves to the simplest continuum model
with the straight-line threshold; dependence of $\xi(\ch\varphi)$
on the threshold curvature was investigated in~\cite{N}.

\begin{figure}[ht]
\begin{center}
\includegraphics[bb=0 350 290 540]{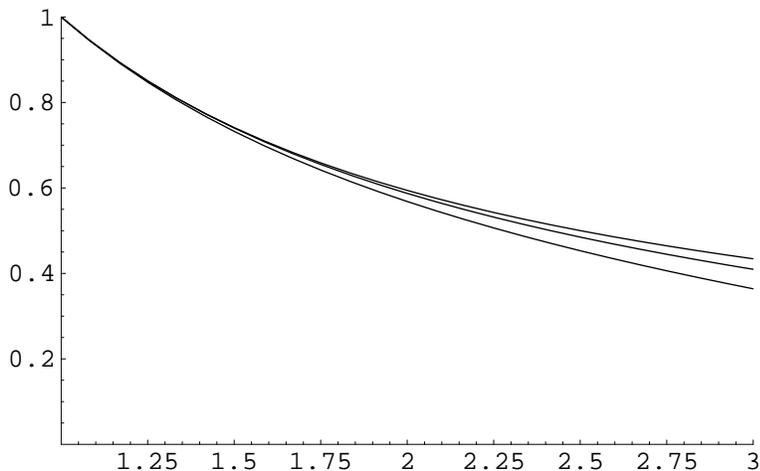}
\end{center}
\caption{Isgur-Wise form factor $\xi(\ch\varphi)$ at Borel parameter values
$E=1.7$, 2, and 2.5}
\label{F2}
\end{figure}

The sum rule for $\xi(\ch\varphi)$ is well known~\cite{R,N}.
It contains two main terms:
the perturbative contribution and the quark condensate one.
The quark condensate contribution is constant
(up to nonlocality effects discussed in~\cite{R}).
The perturbative contribution contains the light quark propagator
from $x_c$ to $x_b$, and falls as $1/(x_b-x_c)^4$.
The Euclidean distances from 0 to $x_b$ and from 0 to $x_c$
are $1/(2E)$; the angle between these lines is $\varphi$.
Hence the distance from $x_b$ to $x_c$ is $\ch\frac{\varphi}{2}/E$.
The perturbative contribution is suppressed as $1/\ch^4\frac{\varphi}{2}$
compared to the 2-point (straight-line) case.
This leads to the decreasing of $\xi(\ch\varphi)$ with increasing $\varphi$.
Graphs for $\xi(\ch\varphi)$ at several Borel parameters
are shown at Fig.~\ref{F2}; the remarkable stability is seen.

\begin{figure}[ht]
\begin{center}
\includegraphics[bb=0 350 290 540]{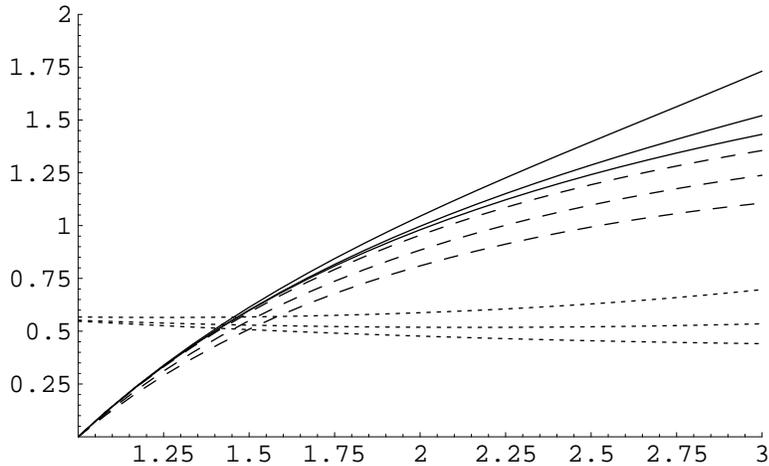}
\end{center}
\caption{$1/m$ correction form factors $\rho_1(\ch\varphi)$ (solid curves),
$\rho_2(\ch\varphi)$ (dashed curves),
and $\rho_4(\ch\varphi)$ (dashed-dotted curves)
at Borel parameter values $E=1.7$, 2, and 2.5}
\label{F3}
\end{figure}

Now we present the new results for the $1/m$ correction form factors
$\rho_{1,2,4}(\ch\varphi)$.
It is seen from the sum rules~(\ref{SR}) that their scale is set by the unit
$k=\left(\frac{\pi^2}{6}|\va{\overline{q}q}|\right)^{1/3}=260$--280MeV.
The dimensionless functions in the sum rules
don't contain large numerical factors.
The numerical analysis of the sum rules confirms this conclusion;
the results are shown at Fig.~\ref{F3}.
The curves $\rho_{1,2}(\ch\varphi)$ are pinned at the origin
by the Luke's theorem;
at $\ch\varphi>1$ they grow and reach the values of order $k$.
On the other hand, $\rho_4(\ch\varphi)$ is small and nearly constant.
Variation of the results with the Borel parameter $E$
allows to estimate their accuracy.

In this work we have investigated the sum rules for $\rho_{1,2,4}(\ch\varphi)$.
It is sufficient for the decay $B\to D$.
Form factors of the decay $B\to D^*$ also contain $\rho_3$.
In order to obtain sum rules for it, it is necessary to consider
an axial correlator in addition to~(\ref{corr}).
Such an analysis will be presented elsewhere.

When this work was completed and presented at the seminar at SLAC,
M.~Neubert informed us about his preprint~\cite{N3}.
In this work the sum rules for the subleading $B\to D^{(*)}$ form factors
were obtained in the framework of HQET.
The results are similar to our ones.
We are grateful to M.~Neubert for giving us the preprint~\cite{N3}
and for the useful discussion.

\appendix

\section{Correlators in QCD}
\label{App}

Here we present the spectral densities and quark condensates' contributions
up to dimension 6 for the correlators~(\ref{corr})
and the correlator similar to $K_{\mu\nu}$ but with the axial current
$\overline{c}\gamma_\mu \gamma_5 b$
(it has the structure $K_A g_{\mu\nu}+K_{++}p_\mu p_\nu+K_{+-}p_\mu q_\nu
+K_{-+}q_\mu p_\nu+K_{--}q_\mu q_\nu$).
The results for all correlators have been produced
by a single REDUCE~\cite{REDUCE} program in which only few lines
with the $\gamma$-matrix structures of currents varied.
this allows to avoid bugs in the programs for separate channels.
To minimize the probability of printing errors, a single REDUCE source
was used both for algebraic checking and for production of \LaTeX{} source
of equations in this appendix (using the package RLFI by R.~Liska
from the REDUCE library~\cite{REDUCE}).

The spectral densities are
\begin{eqnarray*}
\rho_+ &=& \frac{N}{8\pi^2\Delta^{3/2}}
\Bigl[-t(2x_bx_c+(m_b+m_c)a_+)+(m_b-m_c)a_+b\Bigr],
\\
\rho_- &=& \frac{N}{8\pi^2\Delta^{3/2}}
\Bigl[-\bigl((m_b-m_c)^2-t\bigr)(m_b-m_c)a_++2x_bx_c(x_b-x_c)
\\
&&{}+a_-\bigl((3m_c-m_b)x_b+(3m_b-m_c)x_c\bigr)\Bigr],
\\
\rho_V &=& \frac{N}{8\pi^2\Delta^{3/2}}
\Bigl[\bigl((m_b-m_c)^2-t\bigr)a_++(x_b-x_c)a_-\Bigr],
\\
\rho_A &=& \frac{N}{8\pi^2\Delta^{3/2}}
\Bigl[\Delta a_++2m_b\bigl(x_bx_ct+(x_b-x_c)(m_c^2x_b-m_b^2x_c)\bigr)\Bigr],
\\
\rho_{++} &=& \frac{Nm_b}{8\pi^2\Delta^{5/2}}
\Bigl[\Delta t(x_b+x_c)+\Delta\bigl(6x_bx_c-(m_b^2-m_c^2)(x_b-x_c)\bigr)
\\
&&{}+6t(x_b+x_c)(2x_bx_c+m_c^2x_b+m_b^2x_c)-6x_bx_cb^2\Bigr],
\\
\rho_{+-} &=& \frac{N}{8\pi^2\Delta^{5/2}}
\Bigl[-\Delta tx_b(m_b-m_c)
\\
&&{}-\Delta\Bigl((x_b-x_c)a_+-(m_b-m_c)\bigl((m_b^2+m_c^2)x_b-2m_b^2x_c\bigr)
\Bigr)
\\
&&{}-6m_b\bigl(x_bx_ct+(x_b-x_c)(m_c^2x_b-m_b^2x_c)\bigr)b\Bigr],
\\
\rho_{-+} &=& \frac{N}{8\pi^2\Delta^{5/2}}
\Bigl[-\Delta tx_b(m_b+m_c)
\\
&&{}+\Delta\Bigl((x_b-x_c)a_-+(m_b+m_c)\bigl((m_b^2+m_c^2)x_b-2m_b^2x_c\bigr)
\Bigr)
\\
&&{}-6m_b\bigl(x_bx_ct+(x_b-x_c)(m_c^2x_b-m_b^2x_c)\bigr)b\Bigr],
\\
\rho_{--} &=& \frac{Nm_b}{8\pi^2\Delta^{5/2}}
\Bigl[\Delta t(x_b-x_c)
\\
&&{}+\Delta\bigl(-2x_c(2x_b-x_c)-(3m_c^2+m_b^2)x_b+(3m_b^3+m_c^2)x_c\bigr)
\\
&&{}-6t(x_b-x_c)(m_c^2x_b-m_b^2x_c)
\\
&&{}-6\bigl(x_b(x_b-x_c)+(m_b^2+m_c^2)x_b-2m_b^2x_c\bigr)
\\
&&{}\bigl(x_c(x_c-x_b)+(m_c^2+m_b^2)x_c-2m_c^2x_b\bigr)\Bigr],
\end{eqnarray*}
where $\Delta=s_b^2+s_c^2+t^2-2s_b s_c-2s_b t-2s_c t$,
$x_{b,c}=s_{b,c}-m_{b,c}^2$, $a_\pm=m_cx_b\pm m_bx_c$,
$b=x_b-x_c+m_b^2-m_c^2$.

The results for
$\overline{q}(\gamma_5,\gamma_\nu)c\to\overline{q}(1,\gamma_\nu\gamma_5)c$,
$\overline{b}(\gamma_\mu,\gamma_\mu\gamma_5)c\to
\overline{b}(\gamma_\mu\gamma_5,\gamma_\mu)c$
can be easily obtained by
\[ K'_i=-K_i(m_c\to-m_c). \]
There is an interesting check of these formulae.
If we multiply a correlator with $\overline{q}\gamma_\nu c$ by $p_{c\nu}$
and use the identity
$S_q(k)\gamma_\nu S_c(p_c+k)p_{c\nu}=m_cS_q(k)S_c(p_c+k)+S_q(k)-S_c(p_c+k)$,
we obtain the corresponding correlator with $\overline{q}c$
plus terms having no double discontinuity.
Therefore
\begin{eqnarray*}
\rho_A + (s_b+3s_c-t)\rho_{++} + (s_b-s_c-t)\rho_{+-} &=&
-2\rho_+(m_c\to-m_c),
\\
-\rho_A + (s_b+3s_c-t)\rho_{-+} + (s_b-s_c-t)\rho_{--} &=&
-2\rho_-(m_c\to-m_c).
\end{eqnarray*}

\begin{sloppypar}
The quark condensates' contributions are
\begin{eqnarray*}
\frac{K^q_+}{|\va{\overline{q}q}|} &=&
- \frac{m_b+m_c}{2x_bx_c}
+ \frac{m_0^2}{4} \biggl[ (m_b+m_c)
\left(\frac{m_b^2}{x_b^3x_c}+\frac{m_c^2}{x_c^3x_b}+\frac{c}{3x_b^2x_c^2}
\right)
\\
&&{} + \frac23
\left(\frac{2m_b+m_c}{x_b^2x_c}+\frac{2m_c+m_b}{x_c^2x_b}\right)\biggr]
\\
&+& \frac{m_1^2}{3} \biggl[ -(m_b+m_c)a_+
\left(\frac{m_b^2}{x_b^4x_c^2}+\frac{m_c^2}{x_c^4x_b^2}
+\frac23\frac{(m_b-m_c)^2-t}{x_b^3x_c^3}\right)
\\
&&{} + \frac23
\left(\frac{m_b(m_b+2m_c)}{x_b^3x_c}+\frac{m_c(m_c+2m_b)}{x_c^3x_b}
+\frac{c-2t}{x_b^2x_c^2}\right)
\\
&&{} + 4\left(\frac1{x_b^2x_c}+\frac1{x_c^2x_b}\right) \biggr]
\\
\frac{K^q_-}{|\va{\overline{q}q}|} &=&
- \frac{m_b-m_c}{2x_bx_c}
+ \frac{m_0^2}{4} \biggl[ -(m_b-m_c)
\left(\frac{m_b^2}{x_b^3x_c}+\frac{m_c^2}{x_c^3x_b}+\frac{c}{3x_b^2x_c^2}
\right)
\\
&&{} + \frac23
\left(-\frac{3m_b-m_c}{x_b^2x_c}+\frac{3m_c-m_b}{x_c^2x_b}\right)\biggr]
\\
&+& \frac{m_1^2}{3} \biggl[ (m_b-m_c)a_+
\left(\frac{m_b^2}{x_b^4x_c^2}+\frac{m_c^2}{x_c^4x_b^2}
+\frac23\frac{(m_b-m_c)^2-t}{x_b^3x_c^3}\right)
\\
&&{} - \frac43\frac{(m_bx_b+m_cx_c)a_-}{x_b^3x_c^3}
- \frac83\left(\frac1{x_b^2x_c}-\frac1{x_c^2x_b}\right) \biggr]
\\
\frac{K^q_V}{|\va{\overline{q}q}|} &=&
- \frac1{2x_bx_c} + \frac{m_0^2}{4} \left[
\frac{m_b^2}{x_b^3x_c}+\frac{m_c^2}{x_c^3x_b}+\frac{c}{3x_b^2x_c^2}
+\frac{2}{3x_b^2x_c}\right]
\\
&+& \frac{m_1^2}{3} \biggl[ -a_+
\left(\frac{m_b^2}{x_b^4x_c^2}+\frac{m_c^2}{x_c^4x_b^2}
+\frac23\frac{(m_b-m_c)^2-t}{x_b^3x_c^3}\right)
\\
&&{} - \frac43\left(-\frac{m_b}{x_b^3x_c}+\frac{2m_c}{x_c^3x_b}
+\frac{m_b+m_c}{x_b^2x_c^2}\right) \biggr]
\\
\frac{K^q_A}{|\va{\overline{q}q}|} &=&
- \frac{(m_b+m_c)^2-t}{2x_bx_c} - \frac1{2x_b} - \frac1{2x_c}
\\
&+& \frac{m_0^2}{4} \biggl[
\left((m_b+m_c)^2-t\right)
\left(\frac{m_b^2}{x_b^3x_c}+\frac{m_c^2}{x_c^3x_b}+\frac{c}{3x_b^2x_c^2}
\right)
\\
&&{}+\frac{m_b^2}{x_b^3}+\frac{m_c^2}{x_c^3}
+\frac{3m_b^2+4m_c^2+9m_bm_c-4t}{3x_b^2x_c}
\\
&&{}+\frac{2m_b^2+3m_c^2+3m_bm_c-2t}{3x_c^2x_b}
+\frac23\left(\frac1{x_b^2}-\frac1{x_bx_c}\right)\biggr]
\\
&+& \frac{m_1^2}{3} \biggl[ -\left((m_b+m_c)^2-t\right)a_+
\biggl(\frac{m_b^2}{x_b^4x_c^2}+\frac{m_c^2}{x_c^4x_b^2}
\\
&&{}+\frac23\frac{(m_b-m_c)^2-t}{x_b^3x_c^3}\biggr)
-\frac{m_b^3}{x_b^4}-\frac{m_c^3}{x_c^4}
\\
&&{}+\frac{m_b(3m_b^2+2m_c^2+3m_bm_c-2t)}{3x_b^3x_c}
\\
&&{}-\frac{m_c(10m_b^2+11m_c^2+19m_bm_c-10t)}{3x_c^3x_b}
\\
&&{}-\frac{6m_b^3+5m_b^2m_c+7m_bm_c^2-2m_c^3-2(3m_b-m_c)t}{3x_b^2x_c^2}
\\
&&{}+\frac43\left(\frac{m_b}{x_b^3}-\frac{2m_c}{x_c^3}
+\frac{m_b+4m_c}{x_b^2x_c}+\frac{m_b-2m_c}{x_c^2x_b}\right)\biggr]
\\
\frac{K^q_{++}}{|\va{\overline{q}q}|} &=&
\frac1{2x_bx_c} - \frac{m_0^2}{4} \left[
\frac{m_b^2}{x_b^3x_c}+\frac{m_c^2}{x_c^3x_b}
+\frac{m_b^2+2m_c^2-2t}{3x_b^2x_c^2}\right]
\\
&+& \frac{m_1^2}{3} \biggl[ m_b
\left(\frac{m_b^2}{x_b^4x_c}+\frac{m_c^2}{x_c^4x_b}
+\frac{m_b^2+2m_c^2-2t}{3x_b^3x_c^2}+\frac{m_c^2+2m_b^2-2t}{3x_c^3x_b^2}
\right)
\\
&&{} + 2\left(-\frac{m_b}{x_b^3x_c}+\frac{m_c}{x_c^3x_b}
+\frac{m_c}{x_b^2x_c^2}\right) \biggr]
\\
\frac{K^q_{+-}}{|\va{\overline{q}q}|} &=&
- \frac{m_0^2}{12}
\left[\frac{m_b(m_b-m_c)}{x_b^2x_c^2}-\frac{2}{x_b^2x_c}\right]
\\
&-& \frac{m_1^2}{9} \biggl[ (m_b-m_c)
\left(-\frac{m_b^2}{x_b^3x_c^2}+\frac{3m_c^2}{x_c^4x_b}
+2\frac{m_b^2+m_c^2-t}{x_b^3x_c^2}\right)
\\
&&{} + 2\left(\frac{m_b}{x_b^3x_c}-\frac{m_c}{x_c^3x_b}
+\frac{2m_b-5m_c}{x_b^2x_c^2}\right) \biggr]
\\
\frac{K^q_{-+}}{|\va{\overline{q}q}|} &=&
- \frac{m_0^2}{12}
\left[\frac{m_b(m_b+m_c)}{x_b^2x_c^2}-\frac2{x_b^2x_c}+\frac2{x_c^2x_b}\right]
\\
&-& \frac{m_1^2}{9} \biggl[ (m_b+m_c)
\left(-\frac{m_b^2}{x_b^3x_c^2}+\frac{3m_c^2}{x_c^4x_b}
+2\frac{m_b^2+m_c^2-t}{x_b^3x_c^2}\right)
\\
&&{} + 2\frac{(x_b-x_c)a_-}{x_b^3x_c^3}\biggr]
\\
\frac{K^q_{--}}{|\va{\overline{q}q}|} &=&
- \frac1{2x_bx_c} + \frac{m_0^2}{4}
\biggl[\frac{m_b^2}{x_b^3x_c}+\frac{m_c^2}{x_c^3x_b}
+\frac{3m_b^2+2m_c^2-2t}{3x_b^2x_c^2}
\\
&&{}+\frac23\left(\frac2{x_b^2x_c}+\frac1{x_c^2x_b}\right)\biggr]
\\
&+& \frac{m_1^2}{3} \biggl[ m_b
\biggl(-\frac{m_b^2}{x_b^4x_c}+\frac{m_c^2}{x_c^4x_b}
-\frac{3m_b^2+2m_c^2-2t}{3x_b^3x_c^2}
\\
&&{} + \frac{3m_c^2+2m_b^2-2t}{3x_c^3x_b^2}\biggr)
+\frac23\left(\frac{m_b}{x_b^3x_c}-\frac{3m_c}{x_c^3x_b}
+\frac{m_b+3m_c}{x_b^2x_c^2}\right)\biggr]
\end{eqnarray*}
where $x_{b,c}=p_{b,c}^2-m_{b,c}^2$, $c=2m_b^2+2m_c^2-m_bm_c-2t$;
$m_0^2=i\va{\overline{q}gG^a_{\mu\nu}t^a\sigma_{\mu\nu}q}/\va{\overline{q}q}$,
$m_1^3=-\frac14\va{\overline{q}gJ^a_\mu t^a\gamma_\mu q}/\va{\overline{q}q}
=\frac{C_F}{N}\pi\alpha_s\va{\overline{q}q}$ in the factorization approximation
($J^a_\mu=D^{ab}_\nu G^b_{\mu\nu}=g\sum_{q'}\overline{q}'t^a\gamma_\mu q'$,
$C_F=\frac{N^2-1}{2N}$, $N=3$ is the number of colours).
\end{sloppypar}

\end{document}